\documentclass[aps,prb,twocolumn,floatfix,groupedaddress,showpacs]{revtex4}

\usepackage{graphicx,epsf}
\usepackage{dcolumn}
\usepackage{psfrag}

\newcommand{\mub}{$\mu_B$ }
\newcommand{\mubp}{$\mu_B$}
\newcommand{\tc}{T$_c$ }

\newcommand{\zni}{Zn$_i$ }

\newcommand{\mnco}{Zn$_{1-x}$Mn$_x$O }

\newcommand{\znco}{Zn$_{1-x}$Co$_x$O }
\newcommand{\zncop}{Zn$_{1-x}$Co$_x$O}
\newcommand{\znmn}{Zn$_{1-x}$Mn$_x$O }

\newcommand{\zncu}{Zn$_{1-x}$Cu$_x$O }

\begin{document}


\title{Magnetic defects promote ferromagnetism in \znco}

\author{C.H. Patterson}
\affiliation{Quantum Theory Project, PO Box 118435, University of Florida, Gainesville, Florida 32611-8435}
\altaffiliation{Permanent address: School of Physics, University of Dublin, Trinity College, Dublin 2, Ireland.}

\date{\today}

\begin{abstract}
Experimental studies of \znco as thin films or nanocrystals have found ferromagnetism and Curie temperatures above
room temperature and that $p$- or $n$-type doping of \znco can
change its magnetic state. Bulk \znco with  a low defect density and $x$ in the range 
used in experimental thin film studies exhibits ferromagnetism only at 
very low temperatures. Therefore defects in thin film samples or nanocrystals may play an important 
role in promoting magnetic interactions between Co ions in \zncop. The electronic structures of Co substituted for Zn in ZnO,
Zn and O vacancies, 
substituted N and interstitial Zn in ZnO were
calculated using the B3LYP hybrid density functional in a supercell. The B3LYP functional predicts a 
band gap of 3.34 eV for bulk ZnO, close to the experimental value of 3.47 eV.
Occupied minority spin Co 3$d$ levels are 
at the top of the valence band and unoccupied levels lie above the conduction band minimum. Majority spin Co 3$d$ levels hybridize
strongly with bulk ZnO states.
The neutral O vacancy and interstitial Zn are deep and shallow donors, respectively.
The Zn vacancy is a deep acceptor and the acceptor level for substituted N is at mid gap. 
The possibility that $p$- or $n$-type dopants
promote exchange coupling of Co ions was investigated by computing total energies of magnetic states of ZnO supercells containing
two Co ions and an oxygen vacancy, substituted N or interstitial Zn in various charge states. 
The neutral N defect and the singly-positively charged O vacancy are the only defects which strongly promote ferromagnetic
exchange coupling of Co ions at intermediate range.
\end{abstract}


\maketitle

\section{\label{Introduction}Introduction}
Following initial reports of high \tc ferromagnetism in \znco \cite{Ueda01}, there have been contradictory reports on the magnetic state
of this system \cite{Kim02,Lee02,Venkatesan04,Schwartz04,Kittilstved05}.
Electron paramagnetic resonance (EPR) measurements show that it is possible to produce \znco with x $\leq$ 0.1 \cite{Hausmann68,Jedrecy04,
Ozerov05} in which Co is substituted for Zn rather than forming clusters. These studies also show that nanocrystalline \znco is paramagnetic
in this concentration range and at low temperature \cite{Jedrecy04,Ozerov05}; ferromagnetism was observed only at very low temperature.
Another agent must therefore be responsible for promotion of high \tc ferromagnetism in \zncop. 
Very recently it has been shown that high \tc ferromagnetism
can be switched reversibly when \znco or \znmn nanocrystals are capped by O or N
\cite{Kittilstved05,Kittilstved05a} or when \znco thin films are exposed to Zn vapor \cite{Schwartz04}.
The electric conductivity of ZnO changes markedly when it is annealed in vacuum to produce an $n$-type material \cite{Chang01,Tuan04},
while annealing in an oxygen
atmosphere restores the insulating behavior of ZnO.
Venkatesan \textit{et al.} have shown that the magnetic moment per Co ion 
in \znco at room temperature is reduced from 2.8 \mub per Co ion
to zero as the oxygen pressure is varied during annealling \cite{Venkatesan04}.

Here we report results of all-electron B3LYP\cite{Becke93,Stephens94}  hybrid density functional theory (DFT) calculations 
of the electronic states induced by defects (O vacancy, V$_O$, Zn vacancy, V$_{Zn}$,
Zn interstitial, Zn$_i$, and N substituted for O, N$_O$) as well as the electronic structures of bulk ZnO and \znco with $x$ = 0.028.
Several DFT calculations on \znco have been reported recently \cite{Risbud03,Spaldin04,Lee04,Petit04,Sluiter05} and
two of these\cite{Spaldin04,Petit04} have considered the role of defects in promoting ferromagnetism. One of the difficulties in using
DFT to study defects in an insulating material with defect induced electronic states in mid gap 
is severe underestimation of the band gap and overestimation
of exchange coupling constants \cite{Martin97}. In contrast, the B3LYP hybrid functional predicts band gaps for metal oxides which are
in good agreement with experimental values \cite{Cora04} and exchange coupling constants which are also in good 
agreement with experiment \cite{Martin97}. Self-interaction correction of DFT results in an improved prediction for the band gap of ZnO
\cite{Petit04}. \znmn and \zncu have been studied recently using B3LYP hybrid functional calculations \cite{Feng04}.

\begin{figure}
\includegraphics[height=9cm]{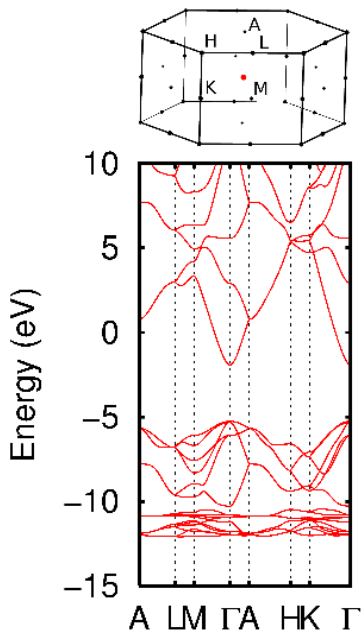}
\caption{\label{fig1}(Color online) Brillouin zone and band structure for bulk ZnO with the wurtzite structure.}
\end{figure}
                                                                                                                 
Defects in ZnO \cite{Muller63} and transition metal doped ZnO \cite{Hausmann68,Anderson67,Hausmann69,Koidl77} 
were originally studied in the 1960's and 1970's by magnetic resonance and optical techniques. 
Recent magnetic resonance \cite{Reynolds98,Carlos01,Leiter01,Gorelkinskii04,Vlasenko05,Tuomisto05}, carrier mobility \cite{Look99} and 
positron annihilation spectroscopy \cite{Tuomisto03} and first principles calculations \cite{Erhart05,Zhang01,Park02}
have given a fuller picture of defect induced electronic states in ZnO.
Magnetic resonance techniques \cite{Ozerov05,Jedrecy04,Diaconu05} have been applied to \znco and \znmn films and nanoclusters.  
As noted above, these studies \cite{Jedrecy04,Diaconu05,Ozerov05,Kobayashi05} as well as X-ray magnetic 
circular dichroism \cite{Tuan04,Kobayashi05}, which probe the local crystal 
structure about the transition metal ion,
have concluded that the transition metal ions substitute for a Zn ion.
Expitaxial growth of thin films with (110) orientation have been achieved \cite{Tuan04} and there 
is considerable anisotropy of the magnetism in the thin film \cite{Venkatesan04}.

The remainder of this paper is organized as follows: Details of calculations are described in Section II; results of B3LYP calculations
on the defects mentioned above, \znco as well as two Co ions in the same unit cell as V$_O$, Zn$_i$ or N$_O$ are described in Section III;
analysis and discussion of these results is given in the final Section.

\section{\label{details}details of calculations}

All-electron B3LYP calculations were performed using the CRYSTAL code \cite{Crystal03} and a wurtzite structure (space group $P6_{3}mc$)
with fundamental lattice constants 3.249 and 5.206 \AA\ parallel to the $a$ and $c$ axes, respectively \cite{Hellwege82}. Defect calculations
were performed using 3x3x2 supercells with dimensions
9.747 and 10.412 \AA\ parallel to the $a$ and $c$ axes.
The band structure of
bulk ZnO was calculated using the wurtzite primitive unit cell and 50 k points in the irreducible Brillouin zone (IBZ) 
and calculations in the 3x3x2 supercell used
13 k points in the IBZ. The Brillouin zone for the primitive unit cell (Fig. \ref{fig1}) and the supercell have the same symmetry, 
hence the same labels are used for high symmetry points in both cases.
All-electron Gaussian orbital basis sets were used for Zn\cite{Jaffe93}, O\cite{Towler94}, Co\cite{Dovesi97} and N\cite{Pandey94}. 
For vacancy defects a basis set corresponding to the ion removed was included at the site of the vacancy ion in order to allow a proper
description of electrons localized in the vacancy. The local crystal structure around defects 
was relaxed and details of their structures are given in the Appendix.
In most cases 18 ion positions, including the position of the vacancy site basis set,
were relaxed for each defect. These are ions within 3.5\AA\ of the defect site.
Combined crystal structure and spin density plots were generated using the XCrySDen package \cite{Kokalj03,xcrysden}.

\section{\label{Results}Results}
\subsection{Bulk ZnO}

\begin{figure}
\includegraphics[height=11cm]{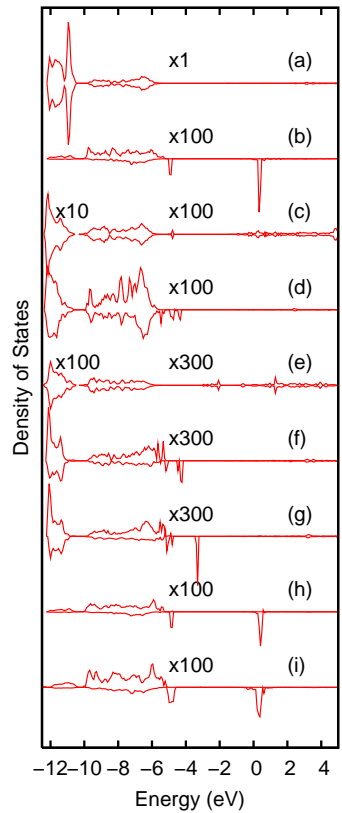}
\caption{\label{fig2}(Color online) Total and atom-projected densities of states for bulk ZnO and ZnO with various defects from 3x3x2 
supercells. Majority spin densities of states are shown immediately above minority spin densities of states. 
(a) Total density of states for ZnO, (b) projected onto Co atom in Co$_{Zn}$, (c) projected onto four Zn ions surrounding
V$_O$, (d) projected onto four O ions surrounding V$_{Zn}$, (e) projected onto six Zn ions surrounding Zn$_i$, (f) projected onto
N ion in N$_O$, (g) projected onto N ion in N$_O$ with two Co ions per unit cell, (h) projected onto two Co ions in N$_O$ with two Co ions
per unit cell, (i) projected onto two Co ions in V$_O$ with two Co ions per unit cell. Scaling factors which have been used are indicated.}
\end{figure}

The band structure of bulk ZnO in the wurtzite phase is shown in Fig. \ref{fig1} and the total bulk density
of states is shown in Fig. \ref{fig2}a.
Band energies have not been shifted to align the valence band maximum (VBM) with any reference energy in order to allow comparison
of the band structure when defects are present or absent. 
The band gap of 3.34 eV obtained using the B3LYP functional is in good agreement 
with the experimental value of 3.47 eV \cite{Madelung91} and band disperison in both valence and conduction bands 
is in good agreement with an earlier plane wave DFT study \cite{Schroer93} and a B3LYP calculation on ZnO\cite{Feng04}.  
The band gap predicted for ZnO by DFT methods is considerably underestimated, making calculations of positions of defect levels difficult 
or impossible;
DFT pseudopotential calculations predict band gaps of 
0.23 eV  \cite{Schroer93}, 0.91 eV \cite{Kohan00} or 0.81 eV \cite{Erhart05} and the gap depends strongly on the particular 
pseudopotential used \cite{Schroer93}.


\subsection{Substituted cobalt Co$_{Zn}$}

\begin{figure}
\includegraphics[height=7cm]{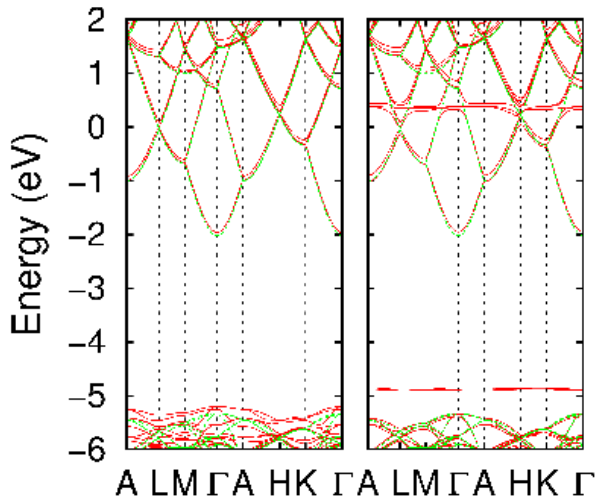}
\caption{\label{fig3}(Color online) Band structure for Co$_{Zn}$ in a 3x3x2 supercell (solid line) and for bulk ZnO in a 3x3x2 supercell 
(dotted line). 
Left panel: majority spin. Right panel: minority spin.}
\end{figure}

Majority and minority spin band structures are shown in Fig. \ref{fig3} and the Co atom-projected density of states 
is shown in Fig. \ref{fig2}b.
Occupied \textit{minority} spin 
Co 3$d$ states with $e$ symmetry are located around 5 eV below the vacuum level,
while unoccupied $t_{2}$ states are located around the vacuum level. 
Majority spin 3$d$ states are fully occupied, hybridize strongly with O 2$p$ states
and do not appear as a sharp feature in the density of states. 
By comparing the bulk ZnO band structure with the Co$_{Zn}$
band structure in Fig. \ref{fig3}, it can be seen that there is a minimal change in the ZnO minority band structure when Co is introduced 
and that the minority
spin 3$d$ states consist of discrete levels. The majority spin conduction band structure of Co$_{Zn}$ is 
nearly identical to that of bulk ZnO but
there is a considerable difference between the majority spin valence band structure for ZnO and Co$_{Zn}$.
Results from resonant photoemission experiments \cite{Wi04,Kobayashi05} show a 
well defined peak at the top of the valence band as well as a broader density of states due to Co extending down to 8 eV below the Fermi level;
the photoemission peak near the Fermi level may therefore arise predominantly from emission from 3$d$ minority spin states.
The large energy splitting of occupied $e$ and unoccupied $t$ minority spin $d$ electrons found here is caused by differences in exchange
energies of these levels and not crystal field splitting; a similar study of \znmn \cite{Feng04}, in which $d$ levels are entirely occupied
or entirely empty, shows a crystal field ($e$ and $t_2$) splitting of less than 1 eV.

Optical spectra of \znco in both earlier single-crystal studies \cite{Anderson67,Koidl77} and 
more recent ferromagnetic thin film studies \cite{Schwartz04}, show 
Co$^{2+}$ $d$-$d$ transitions at mid-gap in ZnO. 
However, optical spectra in materials with highly localized excited states
yield excitation energies for the N-electron state,
but do not provide good estimates of ionization potential and electron affinity differences because the large electron-hole attraction
energy in the optically excited N-electron state
reduces the excitation energy well below the difference in ionization potential and electron affinity.
Photoemission experiments \cite{Wi04,Kobayashi05} and band structure calculations, which involve N+1 and N-1 electron 
energy levels, provide the best estimates of the relevant energy level positions. 
However, observation of photocurrents observed in \znco \cite{Kittilstved05} with a sub-gap conduction threshold may place the unoccupied $d$  
levels within the band gap.

\subsection{Oxygen vacancy V$_O$}

\begin{figure}
\includegraphics[height=7cm]{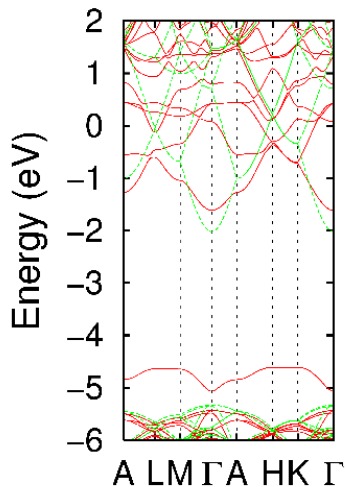}
\caption{\label{fig4}(Color online) Band structure for V$_O$ in a 3x3x2 supercell (solid line) 
and for bulk ZnO in a 3x3x2 supercell (dotted line).}
\end{figure}

B3LYP calculations were performed for one O vacancy in a 3x3x2 supercell. 
The deep defect level associated with the two 'dangling electrons' in a neutral O vacancy is shown in Fig. \ref{fig4} and the densities of
states projected onto the four Zn ions surrounding the vacancy and the vacancy site itself
is shown in Fig. \ref{fig2}c. 
There is no spin polarisation for the neutral vacancy and only a minor change in the ZnO valence band structure.
The position of the V$_O$ and V$_{O}^{+}$ levels have been determined to lie 2.1 and 2.3 eV below the conduction band minimum (CBM) by means of 
deep level nonlinear spectroscopy 
\cite{Gavryushin94}.
The B3LYP calculation on the neutral unit cell places the V$_O$ $\epsilon$(0$\vert$+) level 3.0 eV below the CBM;  
when the calculation was repeated with the unit cell
singly-positively charged, the degenerate V$_{O}^{+}$ vacancy level splits into occupied $\epsilon$(+$\vert$2+) and unoccupied 
levels, with the occupied and unoccupied levels
3.1 and 1.6 eV below the CBM, respectively, at the $\Gamma$ point. These values are somewhat deeper than those obtained from nonlinear
spectroscopy, however, inspection of Fig. \ref{fig4} shows that these levels have their lowest values at the $\Gamma$ point.
(The CRYSTAL program allows charged periodic unit cells to be treated by introducing an effective, 
compensating uniform background charge). 
There is considerable dispersion of the defect level because tails of wave functions associated with the defect induced level extend
beyond the supercell boundaries. Previous DFT calculations \cite{Kohan00,Lee01} on this defect 
have shown that the defect level wave function has 
large amplitude on the shell of O ions outside the shell of Zn ions closest to the vacancy site, which may explain why this level
lies so deep.
Supercell calculations on cells larger than the 3x3x2 cell used are not currently feasible.

\subsection{Zn vacancy V$_{Zn}$}

\begin{figure}
\includegraphics[height=7cm]{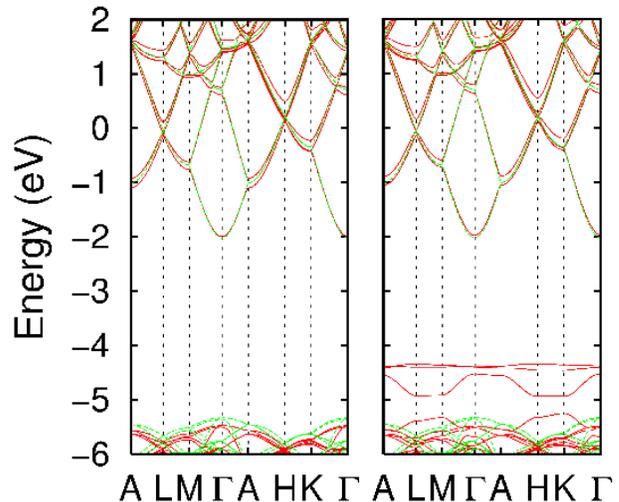}
\caption{\label{fig5}(Color online) Band structure for V$_{Zn}$ in a 3x3x2 supercell (solid line) and for bulk ZnO in a 3x3x2 supercell
(dotted line).
Left panel: majority spin. Right panel: minority spin. 
}
\end{figure}

Zn vacancies are believed to be the dominant deep acceptor states in ZnO \cite{Tuomisto03,Look05}. 
All of the charge states of defects discussed so far, with the exception of the singly-positively charged V$_O$ state, 
have closed shell ground states and no unpaired electrons. 
The neutral Zn vacancy contains two 'dangling holes' so that the minority spin band structure consists of two singly-occupied 
bands and two empty bands (Fig. \ref{fig5}). 
These mainly consist of four O 2$p$ orbitals on the O ions which surround the vacancy site and they
form non-degenerate and quasi-triply-degenerate states with $a$ and $t$ point symmetries. In the neutral V$_{Zn}$ defect
the $a$ state lies just above the VBM and the $t$ state lies above that.  
The degeneracy of the $t$ state is broken to allow occupation of one of 
these orbitals.  
Breaking of the $t$ degeneracy ought to lead to a Jahn-Teller distortion of the tetrahedra of O ions surrounding the vacancy, but there was 
little symmetry breaking when two shells of atoms surrounding the vacancy were relaxed. The two electrons in singly-occupied orbitals 
exist either as a spin singlet (S=0) or triplet (S=1). The wave function form used consists of single determinants for up and down spins
and can only describe the spin triplet state of the vacancy. It is therefore not possible to predict whether the neutral V$_{Zn}$ vacancy will 
have a non-magnetic, singlet or magnetic, triplet ground state on the basis of these calculations.

\subsection{Zn interstitial \zni}

\begin{figure}
\includegraphics[height=7cm]{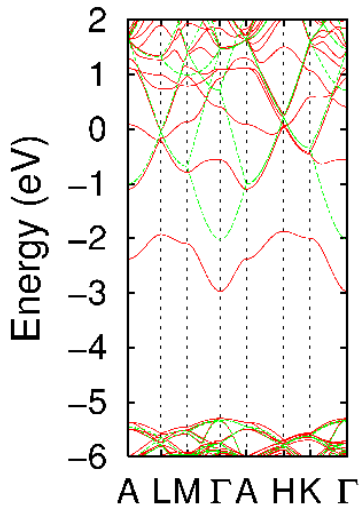}
\caption{\label{fig6}(Color online) Band structure for \zni in a 3x3x2 supercell (solid line) 
and for bulk ZnO in a 3x3x2 supercell (dotted line).}
\end{figure}

Shallow donor states estimated to lie 30 meV below the ZnO CBM have been observed in ZnO subjected to electron irradiation and 
have been assigned to \zni or \zni complexes \cite{Look99}. The energetically favorable site for a single Zn interstitial in ZnO 
is the octahedral 
site \cite{Kohan00}. The band structure for a 3x3x2 ZnO supercell with a single Zn interstitial at the energy minimzed octahedral site
is shown in Fig. \ref{fig6}. A new defect level is formed after hybridization of the 4$s$ level of the Zn interstitial and the lowest 
conduction band (which is predominantly of Zn 4$s$ character). The density of states for the Zn interstitial and the six Zn ions nearest to
it is shown in Fig. \ref{fig2}e. The atom-projected density of states plot shows that
the defect level lies just below the CBM of bulk ZnO (shown in Fig. \ref{fig6} as a
dotted line) 
The defect level disperses by about 1eV throughout the Brillouin zone and the conduction band is considerably modified at low 
energies. In a larger cell with much reduced defect interactions, this strongly dispersive level would likely consist 
of a sharp level within a few
tens of meV of the CBM, 
which would confirm the assignment of the state observed in experiment \cite{Look99}.

\begin{figure}
\includegraphics[width=9cm]{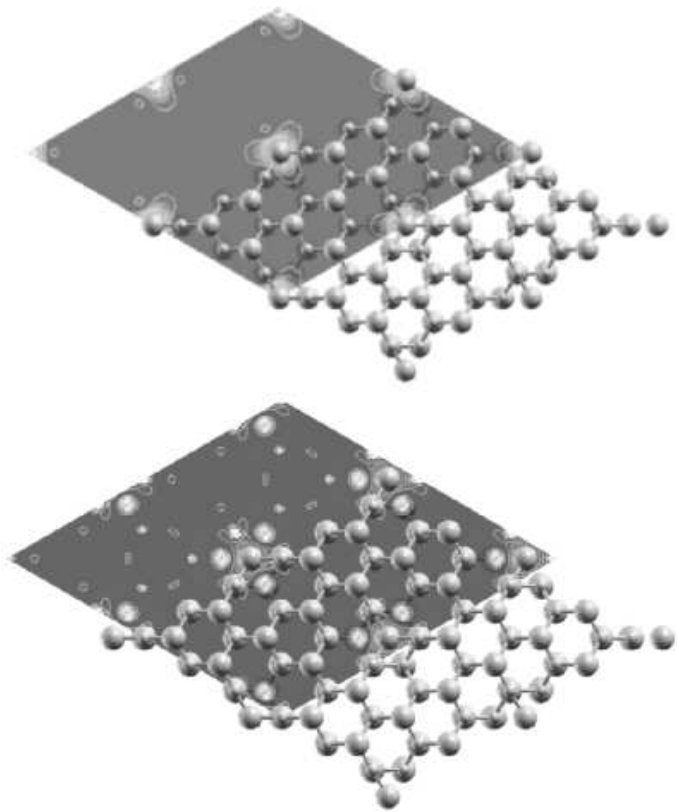}
\caption{\label{fig7}Spin density for \zni in a 3x3x2 supercell with unit positive charge. 
Upper panel: spin density in a plane containing interstitial Zn 
ions. Lower panel: spin density in a plane perpendicular to the $c$ axis containing Zn ions closest to interstitial ions.
O ions are indicated by small grey spheres and Zn ions by large grey spheres.}
\end{figure}

Zn$_i$ in a positive ion state has been proposed as a possible mediator of ferromagnetism in \znco \cite{Schwartz04}. The  
spin density for Zn$_i$ in a 3x3x2 supercell with one electron removed per unit cell 
is shown in Fig. \ref{fig7}. The spin density plot shows that the spin density for the positive ion state is nearly completely localized
on the interstitial ions plus the 12 Zn and O ions surrounding it; this was also found to be the case for the charge density for the 
neutral interstitial associated with the defect level just below the CBM. 
The magnetic moment on the interstitial Zn$_i^+$ ion determined by a Mulliken population analysis 
is 0.37 \mub and the moments on the 6 Zn and 6 O ions immediately
surrounding the vacancy are 0.049 (Zn x3), 0.027 (Zn x3), 0.018 (O x3) and 0.013 (O x3) \mubp, giving a net moment of 0.69 \mub on the 13
ion Zn$_i$ complex and 0.31 \mub on the remaining ions in the supercell.

\subsection{Substituted nitrogen N$_O$}

\begin{figure}
\includegraphics[height=7cm]{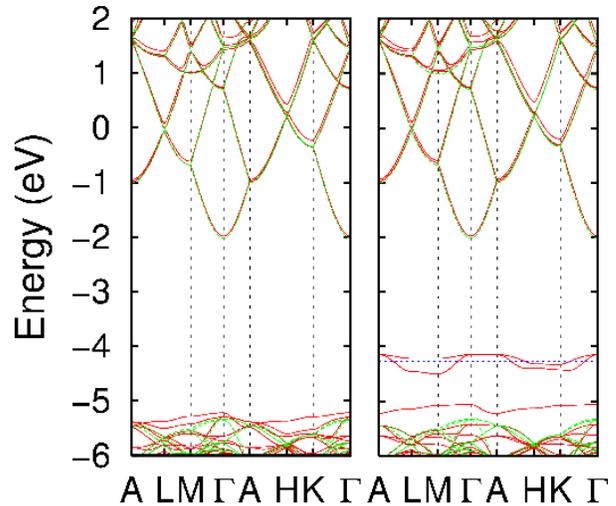}
\caption{\label{fig8}(Color online) Band structure for N$_O$ in a 3x3x2 supercell (solid line) and for bulk ZnO in a 3x3x2 supercell 
(dotted line). 
Left panel: majority spin. Right panel: minority spin. 
The position of the Fermi level is shown as a horizontal dotted line in the minority spin
band structure.}
\end{figure}

Recently it has been shown that N doping can be used to produce $p$-type ZnO
\cite{Look04} and there have been several theoretical studies of co-doped N in \znco and \mnco \cite{Petit04,Sato02}. 
Nitrogen forms a substitutional defect in ZnO \cite{Carlos01} where it leads to introduction of a hole. 
Symmetry constraints allow the hole to be mainly associated with a nearly degenerate pair of N 2$p$
orbitals lying in the $ab$ plane of the wurtzite structure \textit{or} in a non-degenerate 2$p$ orbital parallel to the $c$ axis.
The former orbital occupation was found in the geometry optimized structure for N$_O$ and the semi-metallic band 
structure is shown in Fig. \ref{fig8}, where the partially occupied, weakly dispersive and fully occupied, non-degenerate N 2$p$ states
lie above the ZnO occupied states. 

\begin{figure}
\includegraphics[width=8cm]{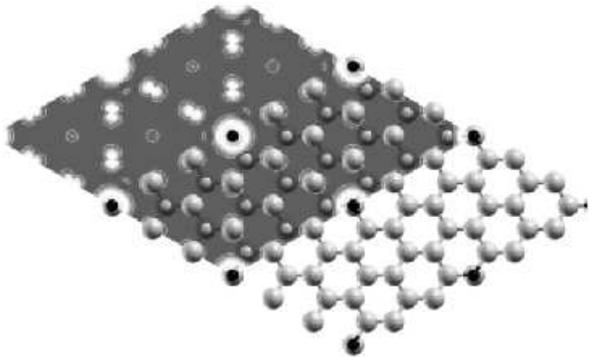}
\caption{\label{fig9}Spin density for N$_O$ in a 3x3x2 supercell in a plane perpendicular to the $c$ axis containing only N and O ions.
O and N ions are indicated by small grey and black spheres, respectively, and Zn ions by large grey spheres.}
\end{figure}

A plot of the spin
density is shown in Fig. \ref{fig9}. The spin density from the N ion delocalizes onto nearby O ions in the same plane
(perpendicular to the $c$ axis) and nearly all spin density is confined to this plane. The magnetic moment on each N ion is 0.66 \mub
and on the O ions with the largest moments, it is 0.03 \mubp.

\subsection{Substituted cobalt with V$_O^{n+}$, N$_O$ or Zn$_i^{n+}$}

\begin{figure}
\includegraphics[width=8cm]{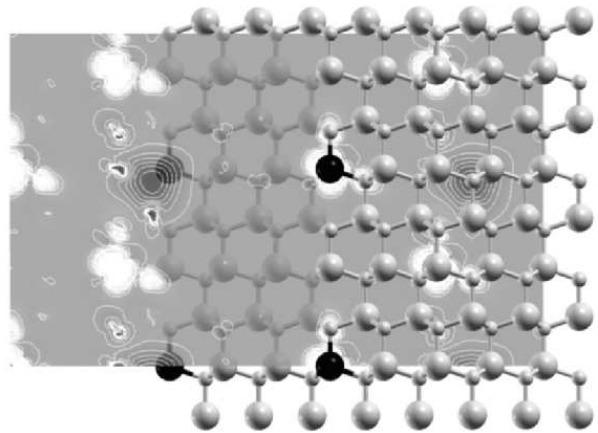}
\caption{\label{fig10}
Spin density for two Co ions and a V$_O^+$ defect in a 3x3x2 supercell. O ions are indicated by small grey spheres 
and Zn and Co ions are indicated by large grey and black spheres, respectively. Spin density on the defect appears as a roughly
spherical density while spin density on Co ions has a mixed $d$ and tetrahedral lobe character.}
\end{figure}

The possibility that neutral or positively charged defects can mediate ferromagnetic interactions in ZnO at intermediate range
(\textit{i.e.} just beyond near neighbor distances) with a strength that could result in Curie temperatures above room temperature was
investigated using total energy calculations. A defect
which mediates ferromagnetism will most likely be magnetic itself as induced magnetic polarisation of the defect is unlikely to result in
strong mediation of ferromagnetism. 
Total energy calculations were performed with two Co ions and one defect per 3x3x2 unit cell and the magnetic moments of the Co ions
parallel or anti-parallel. The minimum distance between 
Co ions in the lattice was 7.66 \AA\ and the Co-V$_O$ and Co-N$_O$ minimum distances were 4.94  and 5.00 \AA. 
The same Zn sites were substituted for Co in Co/V$_O$, Co/N$_O$ and Co/Zn$_i$ calculations.
The same O sites were substituted by V$_O$ or N allowing a direct comparison of the efficacy of magnetic coupling of these
defects in various charge states. In Co/Zn$_i$ calculations the minimum Co-Zn$_i$ distances were 5.10 and 6.68 \AA.
The total energy difference between the two magnetic states
for two Co ions in the absence of a vacancy is less than 1 meV. When a V$_O$ defect was introduced, the total energy differences
for the two magnetic states were less than 1 meV for V$_O$ or V$_O^{2+}$ vacancies and 36 meV for a V$_O^+$ vacancy (with the
ferromagnetic state of the Co ions lower in energy).
The spin density for two Co ions and a V$_O^+$ vacancy with the Co majority spins parallel to each other is shown in Fig. \ref{fig10}
and the density of states projected onto the two Co ions is shown in Fig. \ref{fig2}i.
When a Zn$_i$ interstitial was introduced, the total energy difference was less
than 1 meV for either the neutral or singly-positively charged states of the defect.
When a N$_O$ defect was introduced,
the ferromagnetic configuration of the Co ions was 27 meV lower in energy than the anti-ferromagnetic configuration.

\section{\label{Discussion}Discussion}

\subsection{\label{defects}Single defects}

Recent positron annihilation experiments \cite{Tuomisto05} found that the V$_O$ defect in ZnO subjected to irradiation by high energy electrons
is neutral at low temperature and that these vacancies have a deep donor character with an ionization level 100 meV below the CBM. 
This defect was found to be ten times more abundant than the V$_{Zn}$ vacancy.
Thermodynamic arguments, \cite{Kohan00,Zhang01,Vandewalle01} which take energy levels from DFT calculations corrected for the band gap error,
have shown that V$_O$ is the most thermodynamically stable defect under Zn rich conditions. The positions of the $\epsilon$(0$\vert$+) level
of V$_O$ are predicted to be 2.3 \cite{Kohan00} or 2.4 eV \cite{Zhang01} below the CBM and are therefore in good agreement with the 
value of 2.1 eV determined from nonlinear spectroscopy \cite{Gavryushin94}. This level may also be involved in the 600 nm photoluminescence
(PL) observed in native ZnO by optical detection of electron paramagnetic resonance (ODEPR) \cite{Vlasenko05}; one model for this PL
places the V$_O$ $\epsilon$(0$\vert$+) level 2.5 eV below the CBM \cite{Vlasenko05}.
Using thermodynamic arguments, Kohan and coworkers \cite{Kohan00} and Zhang and coworkers \cite{Zhang01} find that the V$_O$ defect is
a 'negative U' defect and that the $\epsilon$(+$\vert$2+) level lies 0.6 eV below the CBM \cite{Zhang01}; 
in this case the only charge states of the V$_O$ defect which would be observed in equilibrium are the neutral or 
doubly-positively charged states and it would be unlikely that isolated V$_O$ defects would be the main mediators of ferromagnetism in \zncop. 

Positron annihilation studies of electron-irradiated ZnO \cite{Tuomisto03,Tuomisto05} have shown that V$_{Zn}$ vacancies 
exist in a negatively charged state and have given an estimate for the $\epsilon$(2-$\vert$-) level at 2.3 eV below the CBM. Estimates
of the position of this level from DFT calculations \cite{Kohan00} place this level at 2.6 eV below the CBM. We have not performed
calculations on the negatively charged V$_{Zn}$ vacancy, however, the single particle acceptor levels (Fig. \ref{fig5}) lie 2.3 eV below
the CBM, in agreement with both the positron annihilation and DFT results.

The Zn$_i$ interstitial is known to be a shallow donor and estimates \cite{Wagner74,Gavryushin94,Reynolds98,Look98,Look99} 
of the position of the donor level range between 30 meV 
\cite{Look99} and 200 meV \cite{Gavryushin94} below the CBM.
DFT studies, \cite{Kohan00,Zhang01} where corrections have been made for the band gap error, find that this level lies above the CBM so that 
Zn$_i$ is spontaneously doubly ionized. The B3LYP calculations presented here show that the Zn 4$s$ level in Zn$_i$ hybridizes 
strongly with the ZnO conduction band and forms a strongly dispersive band which lies just below the CBM. 
A calculation on a larger supercell (say 5x5x3) would allow a more reliable prediction of the charge state and donor level position of this
defect.

B3LYP calculations for N$_O$ presented in Fig. \ref{fig8} show a semi-metallic electronic structure for this defect with the Fermi level
2.2 eV below the CBM. An earlier DFT study of this system found the acceptor level for N$_O$ to be 0.4 eV above the VBM \cite{Park02}.

\subsection{\label{exchange}Exchange coupling mechanism}

\begin{figure}
\includegraphics[width=4cm]{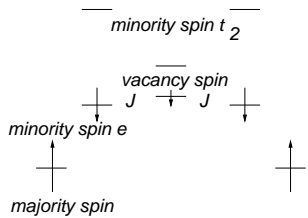}
\caption{\label{fig11}
Energy level diagram for two Co ions and an oxygen vacancy in the ferromagnetic ground state. The exchange integrals which are important
for ferromagnetic exchange coupling are shown as \textit{J} symbols. }
\end{figure}

Substantial ferromagnetic exchange coupling has been found only in two instances in this work and both involve a defect with a magnetic
moment and defect spin densities which overlap with Co ion spin densities. 
The ground state configuration is one where 
majority Co $t_2$ spins are parallel and minority $e$ spins are parallel to each other and the oxygen vacancy spin (Fig. \ref{fig11}), 
so that there are exchange 
couplings between these three spins which lead to an overall ferromagnetic ground state. 
Direct exchange coupling of Co
ions at this separation is less than 1 meV, as demonstrated by total energy calculations when no oxygen vacancy is present. 
When the spin on the defect is removed by a double ionization of the defect the exchange coupling is reduced from 36 to 1 meV, 
demonstrating the importance of the spin of the defect in promoting exchange coupling.
No substantial exchange coupling was found for the positively charged Zn$_i^+$ defect which has spin half also. 
This may be because the spin associated with
the interstitial Zn ion does not overlap the Co ion $e$ wave function to an appreciable extent in the configuration studied.

The exchange coupling mechanism outlined here is essentially the same as the impurity band model of Coey
\textit{et al.} \cite{Coey05} in which there is strong spatial and energetic overlap between an impurity band associated with
a defect and a filled $d$ band. 
A double exchange mechanism of ferromagnetic coupling in \znco and \mnco has been outlined, 
for example by Petit \textit{et al.} \cite{Petit04} or
Kittilstved \textit{et al.} \cite{Kittilstved05,Kittilstved05b}, in which certain charge fluctuations are only permitted for
the ferromagnetic state. These charge fluctuations generally lower the total energy and changes to the energy and wave function
can be estimated using perturbation theory. In the approach used here, energy differences and changes to the wave function
are obtained in the self consistent field calculation. Mulliken spin and charge populations for Co ions and for the O vacancy
and four neighboring Zn ions in both ferromagnetic and antiferromagnetic configurations show that when there is strong ferromagnetic
coupling and Co spins are parallel,
the spin localized on the vacancy decreases by around 0.1 \mub compared to when the spins are antiparallel.  
This shows that there is greater delocalization of the vacancy spin
(and consequently greater overlap with the region where Co $e$ levels are located) when Co spins are \textit{parallel}.

Experiment \cite{Tuomisto05} and DFT calculations \cite{Kohan00,Zhang01} indicate that V$_O$ is the 
most abundant defect in ZnO under zinc rich conditions and that V$_{Zn}$ has an abundance an order of magnitude less
than that of V$_O$ \cite{Tuomisto05}.
Experiment \cite{Venkatesan04} shows that magnetism of \znco in $n$-type ZnO is destroyed by annealing in oxygen. 
The calculations presented here show that V$_O$ must be in a singly-positively charged state in order for it to act as an effective
promotor of ferromagnetism. 
Thermodynamic arguments based on DFT calculations \cite{Kohan00,Zhang01} have been made to suggest that V$_O$ is a 'negative U' defect
in which a pair of V$_O^+$ defects will spontaneously transform into V$_O$ + V$_O^{2+}$. Obviously V$_O^+$ must be stable if indeed
a magnetic defect is necessary for stabilization of the ferromegnetic ground state in \zncop.
In $n$-type \znco with an excess of Zn we therefore expect to find both V$_O$ and V$_{Zn}$ defects with more abundance of V$_O$.
If some of these transfer a single electron to V$_{Zn}$ or another acceptor defect then the magnetic defects necessary for
strong ferromagnetic coupling of Co ions may exist.

\subsection{\label{summary}Summary}

The electronic structures of wurtzite ZnO with Co substituted for Zn, oxygen and zinc vacancies, a zinc interstitial and nitrogen substituted 
for oxygen have been computed using the B3LYP hybrid density functional in 3x3x2 supercells. 
Total energy calculations have been performed for two Co ions in a 3x3x2 supercell with an oxygen vacancy, zinc interstitial and substituted
nitrogen. Neutral and positively charged states of the oxygen vacancy and zinc interstitial were studied. Only the singly-positively 
charged oxygen vacancy and neutral substituted N defects resulted in a substantial exchange coupling of two Co ions at intermediate
range ($>$ 7 \AA).

\begin{acknowledgments}
This work was supported by the Irish Higher Education Authority under the PRTLI-IITAC2 program. 
The author wishes to acknowledge discussions with J.M.D. Coey, M. Venkatesan and D.R. Gamelin and the hospitality of R.J. Bartlett at the University 
of Florida.

\end{acknowledgments}

\appendix

\section{structure optimization}
\label{app}

Minimization of the total energy of the Co$_{Zn}$ defect, by varying positions of ions which lie within 3.5 \AA\ of the Co$^{2+}$ ion, results 
in changes in Co-O bond lengths to 1.98 \AA\ parallel to the $c$ axis and 1.97\AA\ in the $ab$ plane. This corresponds to an expansion
of the Zn-O bond parallel to the $c$ axis by 0.03 \AA\ and a slight contraction of bonds nearly in the $ab$ plane by 0.01 \AA.

When positions of ions within 3.5 \AA\ of the V$_O$ vacancy site were varied to minimize the total energy, 
the Zn ions immediately neighboring the vacancy site
relaxed inwards by ~0.25 \AA\ for the Zn-O bond parallel to the $c$ axis and by ~0.14 \AA\ for Zn-O bonds nearly parallel to the $ab$ plane
so that the nearest Zn-Zn distances for these ions were
reduced from 3.21 and 3.25 \AA\ to 2.89 and 2.97 \AA, respectively. These relaxed bond distances are several percent shorter than those
obtained previously in a DFT pseudopotential calculation which used a plane wave basis set \cite{Kohan00} but are close to the distances
found for the neutral oxygen vacancy in ZnO in a more recent plane wave DFT study \cite{Erhart05}.

O ions around the Zn vacancy relax outwards by 0.14 \AA\ for the Zn-O bond parallel to the $c$ axis and 
by 0.16 \AA\ for the remaining three O ions nearest to the vacancy site. 

The 12 Zn and O ions closest to the interstitial Zn, as well as the interstitial itself, were relaxed for the neutral Zn$_i$ vacancy 
calculations described above. 
The optimized Zn$_i$-Zn distances were 2.36 and 2.47 \AA\ and the Zn$_i$-O distances were 2.06 and 2.68 \AA.  

Ionic radii of O and N ions are similar and so only a small distortion of the ZnO lattice is found when the structure of the
N$_O$ defect is allowed to relax. The Zn-N bond lengths parallel to the $c$ axis and nearly parallel to the $ab$ plane, are 1.94(6) \AA\ 
and 1.97(6) \AA, respectively which compares to Zn-O bond lengths of 1.952 and 1.985 \AA\ in the bulk ZnO.



\bibliography{paper}

\end{document}